\documentclass[]{aastex631}

\shorttitle{Mrk 1018 VLBI Proper Motion}
\shortauthors{Walsh et al.}

\graphicspath{{./}{figures/}}

\begin{document}

\title{A VLBI Proper Motion Analysis of the Recoiling Supermassive Black Hole Candidate Mrk 1018}

\author[0000-0003-1551-1340]{Gregory Walsh}
\altaffiliation{NASA West Virginia Space Grant Consortium Graduate Research Fellow}
\affiliation{Department of Physics and Astronomy, 
West Virginia University, 
Morgantown, WV 26506, USA}
\affiliation{Center for Gravitational Waves and Cosmology,
West Virginia University, 
Chestnut Ridge Research Building, 
Morgantown, WV 26506, USA}

\author[0000-0003-4052-7838]{Sarah Burke-Spolaor}
\affiliation{Department of Physics and Astronomy, 
West Virginia University, 
Morgantown, WV 26506, USA}
\affiliation{Center for Gravitational Waves and Cosmology,
West Virginia University, 
Chestnut Ridge Research Building, 
Morgantown, WV 26506, USA}

\author{T. Joseph W. Lazio}
\affiliation{Jet Propulsion Laboratory,
California Institute of Technology,
4800 Oak Grove Drive, Pasadena, CA 91109, USA}

\correspondingauthor{Gregory Walsh}
\email{gvw0001@mix.wvu.edu}

\begin{abstract}
Mrk 1018 is a nearby changing-look AGN that has oscillated between spectral Type 1.9 and Type 1 over a period of 40 years. Recently, a recoiling supermassive black hole (rSMBH) scenario has been proposed to explain the spectral and flux variability observed in this AGN. Detections of rSMBHs are important for understanding the processes by which SMBH binaries merge and how rSMBHs influence their galactic environment through feedback mechanisms. However, conclusive identification of any rSMBHs has remained elusive to date. In this paper, we present an analysis of 6.5 years of multi-frequency Very Long Baseline Array (VLBA) monitoring of Mrk 1018. We find that the radio emission is compact down to 2.4 pc, and displays flux density and spectral variability over the length of our campaign, typical of a flat spectrum radio core. We observe proper motion in RA of the radio core at -36.4 $\pm$ 8.6 $\mu$as yr$^{-1}$ (4.2$\sigma$), or $0.10c \pm 0.02c$ at the redshift of Mrk 1018. No significant proper motion is found in DEC (31.3 $\pm$ 25.1 $\mu$as yr$^{-1}$). We discuss possible physical mechanisms driving the proper motion, including a rSMBH. We conclude that the apparent velocity we observe of the VLBI radio core is too high to reconcile with theoretical predictions of rSMBH velocities and that the proper motion is most likely dominated by an unresolved, outflowing jet component. Future observations may yet reveal the true nature of Mrk 1018. However, our observations are not able to confirm it as a true rSMBH.

\end{abstract}


\section{Introduction} \label{sec:intro}
Supermassive black holes (SMBHs) are thought to reside in the stellar core of most massive galaxies \citep[][]{Kormendy&Ho_13}. These SMBHs are inevitably displaced from their rest positions when massive galaxies interact through major merger events. Over the lifetime of the galaxy merger, the resident SMBHs will lose energy and momentum through dynamical friction, which drives them towards the minimum of the gravitational potential of the newly merged galaxy \citep{begelman+80}. At separations $\lesssim 10$ pc, the SMBHs will become gravitationally bound and form a supermassive black hole binary (SMBHB).
Once the orbital separation of the binary constituents becomes $\lesssim 0.1$ pc, the binary will evolve rapidly due to the emission of low-frequency gravitational waves (GWs), which are detectable by pulsar timing arrays (PTAs) such as the North American Nanohertz Observatory for Gravitational Waves (NANOGrav; \citealp{Witt+20,ngrv_gwb+20,ngrv_limits+21}).  

SMBHBs in which the constituents have misaligned BH spins and/or an unequal mass ratio will have anisotropic emission of gravitational waves \citep[][]{Peres_62,Fitchett&Detweiler_84}. Simulations show that upon coalescence the anisotropic emission can impart a velocity of up to $5000\, \rm{km s^{-1}}$ onto the newly merged SMBH \citep[][]{Lousto&Zlochower_11}. In the most extreme cases, the SMBH could be ejected from its host galaxy's gravitational potential \citep[][]{Lousto+12}.
However, most recoiling SMBHs will remain bound to their host and will undergo oscillations in the host's gravitational potential well for $10^6$ to $10^9$ years \citep{madau&quataert_04,Blecha+11}. This is the so-called recoiling SMBH (rSMBH). 
The rSMBH will carry its broad-line region with it, continuing to accrete and be active as an AGN \citep[][]{madau&quataert_04,blecha&loeb08,Fujita_09}. During this time, the SMBH may be observed spatially offset from the galactic center, and/or through velocity offsets of the broad emission lines, which trace the line-of-sight velocity of the recoiling SMBH, with respect to the narrow emission lines, which trace the galaxy rest-frame velocity.

\object{Mrk 1018} is a nearby ($z=0.042436$), known changing-look AGN \citep{cohen+86} hosted in a galaxy merger, as evidenced by the tidal tails present in the galaxy's photometry (see Figure 1 of \citet{kim+18}). Mrk 1018 has changed its spectral class from Type 1.9 to Type 1 and back to Type 1.9 over a period of 40 years \citep{osterbrock_81,cohen+86,mcelroy+16}.
Modelling of the flux variations observed in optical and X-ray data has suggested that the changing-look nature of Mrk 1018 is due to accretion variability of the AGN \citep{husemann+16,lamassa+17,noda+18,lyu+21}. Recently, \citet[]{kim+18} (hereafter K18) examined archival data of Mrk 1018 to investigate the physical mechanism behind its changing-look nature. Through decomposition of the broad H$\alpha$ emission, K18 found two kinematically distinct red-and blue-shifted components which varied systematically over the 40-year span covered by the archival data.
They interpret this variability as the manifestation of a recoiling SMBH that carries with it two kinematically distinct broad-line regions, which originated from the two precursor SMBHs. 
The changes in spectral class are caused by perturbations of the accretion disk caused by strong tidal forces as the rSMBH passes through pericenter, i.\,e., the galactic dynamical center, in its orbit. 
Using a Bayesian Markov Chain Monte Carlo simulation, they found the candidate rSMBH to be on a highly eccentric orbit with an orbital period of 29.2 years. If confirmed as a rSMBH, this presents one method to reproduce AGN variability seen on short timescales, e.\,g., the changing-look nature of some AGN.

For the first time, we will directly search for evidence of a recoiling SMBH using high-precision astrometry performed with VLBI. As part of a larger study targeting the evolution of SMBHBs in a sample of post-merger galaxies, we have obtained multi-epoch, multi-frequency observations of Mrk 1018 using the Very Long Baseline Array (VLBA), spanning 6.5 years. The paper outline is as follows. Section \ref{sec:data} describes the calibration procedures for the VLBA and Karl G. Jansky Very Large Array (VLA) observations presented in this paper. Section \ref{sec:analysis} presents analyses on the kpc- and pc-scale emission properties, including the proper motion of the VLBI radio source. In Section \ref{sec:discussion} we discuss the physical mechanisms which may be driving the observed proper motion, including the rSMBH hypothesis. Our conclusions are summarized in Section \ref{sec:conclusions}.

\section{Observations and Data Reduction} \label{sec:data}

\subsection{Very Large Array} \label{sec:vla_obs}
We obtained S- (2-4 GHz) and X-band (8-12 GHz) observations of Mrk 1018 with the Karl G. Jansky Very Large Array (VLA; IDs: 19A-472, 20B-298; PI: P. Breiding) between November 15, 2020 and December 12, 2020. The S-band observations had 16 spectral windows, each with 128 MHz bandwidth split into 64 channels of 2 MHz width, continuously covering the frequency range of 2-4 GHz, and were observed with full polarization. The X-band observations had 32 spectral windows, each with 128 MHz bandwidth split into 64 channels of 2 MHz width, continuously covering the frequency range of 8-12 GHz, and were observed with full polarization. The observations were carried out in BnA- and A-configurations to search for sub-kpc-scale radio emission as well as localize the compact radio core with high astrometric precision. 

The observations used \object{3C 138} for bandpass calibration and \object[WISE J021748.93+014449.9]{J0217+0144} to perform phase referencing, and were calibrated non-interactively using the NRAO VLA Pipeline\footnote{https://science.nrao.edu/facilities/vla/data-processing/pipeline}. The calibrated data were flagged, concatenated, and imaged in the Common Astronomical Software Applications (\verb|CASA|; \citealp{CASA_22}) package following standard routines. We produced images at S- and X-band using the \verb|CASA| task \verb|tclean| with Briggs weighting with a robust parameter of 0.7, 10,000 clean iterations and a threshold of 0.03 mJy. To account for the large fractional bandwidths of the S- and X-band observations, $\sim$ 60\% and $\sim$ 40\% respectively, we used Multi-Taylor Multi-Frequency Synthesis (MTMFS; \citealp{MTMFS_11}) deconvolution to produce our images and in-band spectral index maps.  
Our S-band image used a Gaussian beam of size $1.3\arcsec\times0.83\arcsec$, and X-band used a Gaussian beam of size $0.46\arcsec \times 0.21\arcsec$.
To create a two-point spectral index map, we truncated the 8-12 GHz data to match the 2-4 GHz in \textit{uv}-coverage. This deletes the longer baselines at the higher frequencies, allowing for better flux recovery. We then smoothed the 10 GHz synthesized beam to match the 3 GHz synthesized beam using \verb|imsmooth|.
Finally, we used the task \verb|immath| with mode \verb|spix| to create the full spectral index map. 

\subsection{Very Long Baseline Array} \label{sec:vlba_obs}
We obtained S- (2.204-2.460 GHz), C- (4.128-4.384 and 6.428-6.684 GHz), and X-band (8.688-8.944 GHz) observations of Mrk 1018 with the Very Long Baseline Array (VLBA) in two observing programs, spanning 35 observational epochs from August 2, 2014 to June 21, 2016 (ID: BS237), totaling approximately 10.7 hours on-source at S- and C-band, and 9.5 hours on-source at X-band, and 36 observational epochs from June 17, 2020 to February 2, 2021 (ID: BS280), totaling approximately 50 hours on-source at each observing frequency. 
Our observations utilized the dichroic receiver at the VLBA stations to obtain simultaneous observations at 2.3 and 8.8 GHz.
Our C-band observations were split into two central frequencies at 4.3 and 6.6 GHz. Each observing frequency had a single polarization (RR) and 256 MHz of total bandwidth split into 8 spectral windows, each with 128 channels of 250 kHz width. 

Our observations were designed as a VLBA filler program. As such, we were unsure as to which bandpass calibrators would be observable during each observational epoch. We selected and used 3 unique bandpass calibrators, \object[WISE J015218.05+220707.7]{J0152+2207}, \object[4C +15.05]{J0204+1514} and \object[PKS 0235+164]{J0238+1636}, during the duration of our observing campaigns, with each epoch beginning and ending with scans on two of these. They were selected based off of their compactness, high flux densities, and proximity to Mrk 1018 on the sky. We also employed the usage of two phase reference calibrators for our observing and calibration procedures. Information on the bandpass and phase reference calibrators used for Mrk 1018 is provided in Table \ref{tab:vlba_cal}. 
\begin{deluxetable*}{cccccccc}[t!]
\tablewidth{0pt}
\tablecaption{Calibrators for VLBA Observations of Mrk 1018}
\label{tab:vlba_cal}
\tablecolumns{8}
\tablehead{
\colhead{\rm{J2000 Name}} & \colhead{\rm{RA (J2000)}} & \colhead{$\sigma_\mathrm{RA}$ (mas)} & \colhead{\rm{DEC (J2000)}} & \colhead{$\sigma_\mathrm{DEC}$ (mas)} & \colhead{$S_\mathrm{2.4\, GHz}$ (Jy)} & \colhead{\rm{Cal Type}} & \colhead{\rm{Offset} (\degr)} \\
\colhead{(1)} & \colhead{(2)} &\colhead{(3)} & \colhead{(4)} & \colhead{(5)} & \colhead{(6)} & \colhead{(7)} &\colhead{(8)}
}
\startdata
J0152+2207 & 01:52:18.059042 & 0.04 & +22:07:07.69979 & 0.04 & 0.911 & Amp, MPC & 22.67 \\
J0204+1514 & 02:04:50.413893 & 0.03 & +15:14:11.04361 & 0.03 & 1.876 & Amp, MPC & 15.53 \\
J0238+1636 & 02:38:38.930104 & 0.03 & +16:36:59.27455 & 0.03 & 1.102 & Amp, MPC & 18.70 \\
J0208-0047 & 02:08:26.345908 & 0.11 & -00:47:44.29430 & 0.20 & 0.350 & Phase Ref (C1) & 0.74 \\
J0157+0011 & 01:57:10.534897 & 0.12 & +00:11:24.48457 & 0.26 & 0.192 & Phase Ref (C2) & 2.30 \\
\enddata
\tablecomments{
Column 1: Source name.
Column 2: Source Right Ascension (J2000) in hh:mm:ss. All coordinates and their errors were obtained from the ICRF3 \citep[][]{charlot+20}.
Column 3: Right Ascension error in milliarcsec.
Column 4: Source Declination (J2000) in \degr:\arcmin:\arcsec.
Column 5: Declination error in milliarcsec.
Column 6: 2.4 GHz flux density reported by \citet{charlot+20}.
Column 6: Calibrator type (amplitude and manual phase calibration or phase reference calibrator).
Column 7: Angular difference in degrees between calibrator and Mrk 1018.
}
\end{deluxetable*}

We calibrated the data in the Astronomical Imaging Processing Software (\verb|AIPS|).
Before any calibration was done, we flagged the 20 frequency channels at the low- and high-frequency edges of each spectral window. Spectral windows 1, 5, and 8 for all S-band observations were removed due to the consistent, dominating presence of radio frequency interference (RFI) during the observations.
Earth orientation parameters were fixed using the \verb|AIPS| procedure \verb|VLBAEOPS| and ionospheric delays were calibrated for using \verb|VLBATECR|. As mentioned previously, we selected two unique bandpass calibrators for observing scans for each epoch. These bandpass calibrators were used for manual phase calibration and to solve for bandpass shape. After applying the bandpass correction, we corrected the phases for parallactic angles using \verb|VLBAPANG|.

\begin{figure}[t!]
    \centering
    \includegraphics[scale=0.7]{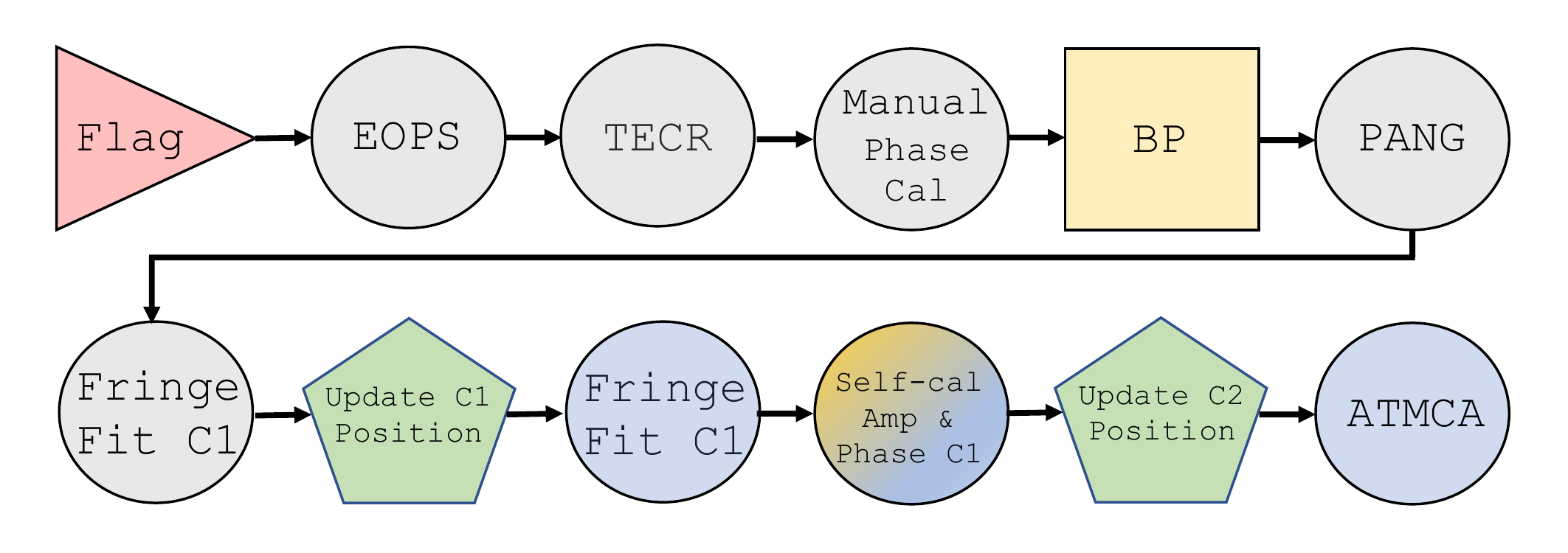}
    \caption{Flowchart of the calibration procedure for each observation. Procedures in red/triangle indicate manual flagging; grey/circle indicate phase calibration; yellow/squares indicate bandpass calibration; green/pentagon indicate an update to the correlator model; light blue circles are phase calibration procedures after correcting for the position of C1. During each step, the latest calibration, flag, and/or bandpass tables are applied. After the initial fringe fitting of the primary phase calibrator, C1, its position is updated in its correlator model to its position given in the ICRF3 \citep[][]{charlot+20}. This assures us that the change of correlation position of C1 due to the adoption of the ICRF3 in between our observing campaigns is not affecting the position of Mrk 1018. After this, C1 is once again fringe fit, and the solutions are interpolated to C1, C2, and Mrk 1018. The procedures after the second fringe fitting of C1 follow those described in the AIPS MEMO 111 (see footnote in Sect~\ref{sec:vlba_obs})}.
    \label{fig:cal_proc}
\end{figure}

Our observational scheme utilized two nearby calibrators, which we denote as C1 and C2, to employ the phase referencing technique. Utilization of two phase reference calibrators will better model atmospheric delays to the target phase, providing higher astrometric precision \citep[][]{Fomalont+05} than the typical single phase reference calibrator. All scans were set up with a C1-C2-T-C1-C2 cadence cycle, with 30-second scans on each phase calibrator and 4 minutes on each target (T) scan. Over the duration of our observing campaigns, each observation spent approximately 73\% on-source and 27\% on calibrator scans. 
\object[PKS 0205-010]{J0208-0047} (C1) was selected as the primary phase calibrator, for which group delay and phase rate calibration solutions were found, 
because of its proximity to Mrk 1018 (0.74\degr).
The correlation position of C1 differed between programs BS237 and BS280 due to the adoption of the ICRF3 on January 1, 2019. Because our observing scripts sourced the ICRF catalogs for C1's position, and phase referencing sets the coordinate reference for the weak target source, this difference in correlation position between our two programs would create false proper motion in the VLBI monitoring of any radio emission associated with Mrk 1018. To account for this, for each observing frequency of epoch in BS237, the position of C1 was shifted by 0.12 mas in RA and -1.3 mas in DEC using the \verb|AIPS| task \verb|CLCOR| to match its ICRF3 position. We discuss this further in Sec.~\ref{sec:systematics}.
Fringe fitting was performed once again on C1 after updating its correlator model and these solutions were interpolated to itself, \object[WISEA J015710.53+001124.1]{J0157+0011} (C2) and Mrk 1018.

For the C- and X-band data, we employed the \verb|AIPS| task \verb|ATMCA| to perform multi-calibrator phase referencing. Due to the high fraction of flagging in the 2.3 GHz observations, \verb|ATMCA| could not find good solutions to the residual phase delays at this frequency. As such, the 2.3 GHz data were calibrated only using standard phase referencing techniques. A description of the \verb|ATMCA| calibration method is described in the AIPS MEMO 111\footnote{http://www.aips.nrao.edu/TEXT/PUBL/AIPSMEM111.PS}, however we outline the procedure in the following description for clarity.

After the second run of fringe fitting for the primary phase reference calibrator, C1 was imaged. Using this image as a model, we performed standard self-calibration iterations to account for phase delays due to source structure of C1.
These solutions were interpolated to C2 and Mrk 1018. C2 was then imaged and its position was determined by Gaussian fitting. The difference between the fitted position of C2 and the observational phase center was calculated, and C2 was adjusted to the phase center, if necessary. Phase calibration solutions were determined for both C1 and C2, with the remaining phase delay residuals now solely associated with instrumental phase errors introduced by the differential atmosphere at the pointing location of each antenna. The \verb|AIPS| task \verb|ATMCA| was used to calibrate for these phase delays. The general calibration procedure is outlined in Fig.~\ref{fig:cal_proc}.

To monitor for any positional or flux variability over the duration of our observing campaign, we produced images at a reference frequencies of 2.3 (S-band), 5.4 (C-band), and 8.8 (X-band) GHz for the BS237 and BS280 observing programs separately.
We made each image interactively using CASA's \verb|tclean| task with natural weighting, and each are $1\arcsec\times1\arcsec$ in size. 
For each observing program, the RMS noise in the images produced at 5.4 GHz and 8.8 GHz were within 10\% of the expected thermal noise limit, after taking into account the variable antenna availability and flagging which occurred during each observational epoch.
The RMS noise in the 2.3 GHz images for both BS237 and BS280 were within a factor of $1.5\times$ the expected thermal noise limit. This is not unexpected, as the RFI environment in this frequency band can severely affect the quality of the data. As mentioned previously, at least 3 IFs within this band were removed prior to calibration due to the persistent, dominating presence of RFI during both observing programs.

\section{Analysis}
\label{sec:analysis}
\subsection{Physical Properties of Radio Component}
\label{sec:properties}
\begin{figure}
    \centering
    \includegraphics[scale=0.7]{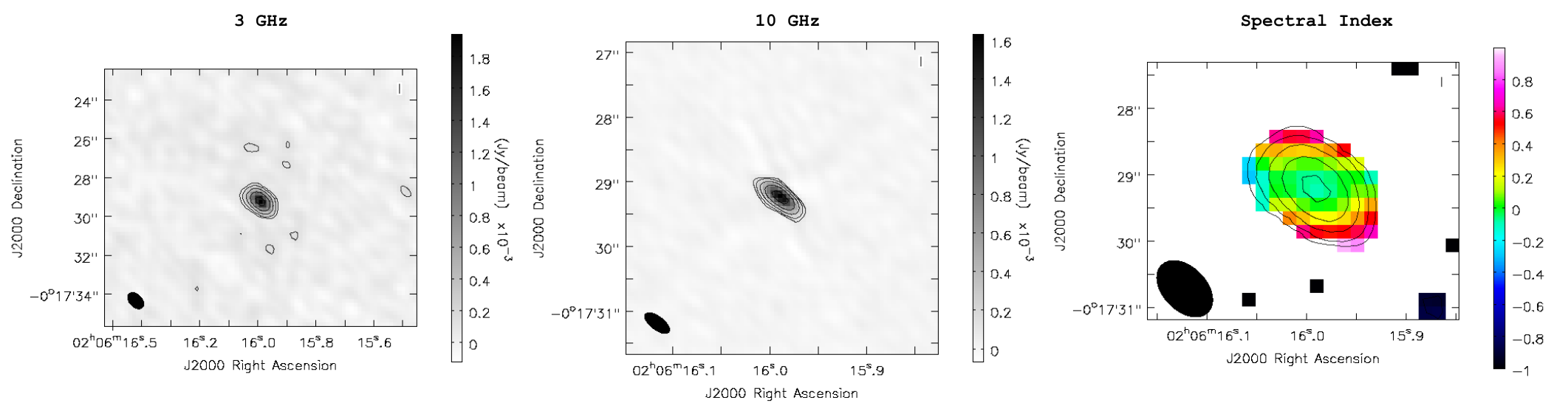}
    \caption{3 GHz (left) and 10 GHz (center) VLA intensity maps of the radio source associated with Mrk 1018. Radio contours are -3, 3, 5, 10, 20, 40 times the image RMS (38 and 30 $\mu$Jy bm$^{-1}$, respectively). The source is unresolved in both images. \textit{Right: Two-point spectral index map of Mrk 1018. This map was produced by \textit{uv}-tapering and resolution-matching our 3 and 10 GHz VLA images. Contours are identical to those presented in the 3 GHz VLA image. The emission has a flat spectral index, $\alpha = -0.13 \pm 0.11$ ($S_\nu \propto \nu^\alpha$) produced by synchrotron self-absorption. This, in conjunction with the compactness of the source, is taken as evidence that this radio emission is associated with a radio core.} The restoring beam is drawn in the lower left corner of each image. The X-shaped structure seen in the 10 GHz is due to residual calibration errors caused by the poor \textit{uv}-coverage of the snapshot observation.}
    \label{fig:vla_intensity}
\end{figure}

Fig.~\ref{fig:vla_intensity} shows the 3 and 10 GHz total intensity maps of the VLA of Mrk 1018. In each of the images, the radio emission is unresolved at the sub-kpc scales probed by our VLA observations: $<$ 600 pc and $<$ 200 pc for the 3 and 10 GHz images, respectively. To determine the flux density at each central frequency, we performed 2D elliptical Gaussian fitting to the unresolved radio component in each image. The error associated with each flux density measurement was found by summing in quadrature the image RMS noise and a standard flux calibration error of 3\% of the peak pixel value \citep{Perley&Butler_17}. The flux densities, their associated errors, the restoring beams, and other properties, are listed in Table \ref{tab:vla_obs}. 
\begin{deluxetable*}{ccccccc}[t!]
\tablewidth{0pt}
\tablecaption{Summary of VLA Observations and Images of Mrk 1018}
\label{tab:vla_obs}
\tablecolumns{7}
\tablehead{
\colhead{$\nu$ (GHz)} & \colhead{$\Delta \nu$ (GHz)} & \colhead{$S_\nu$ (mJy)} & \colhead{$B_{\rm maj}$ ($\arcsec$)} &
\colhead{$B_{\rm min}$ ($\arcsec$)} & \colhead{$PA$ ($\degr$)} & \colhead{$\sigma_{\rm RMS}$ ($\mu$Jy beam$^{-1}$)} \\
\colhead{(1)} & \colhead{(2)} &\colhead{(3)} & \colhead{(4)} & \colhead{(5)} & \colhead{(6)} & \colhead{(7)}
}
\startdata
3.0 & 2 & 1.913 $\pm$ 0.07 & 1.03 & 0.68 & 46 & 38 \\
10.0 & 4 & 1.625 $\pm$ 0.06 & 0.46 & 0.21 & 53 & 30 \\
\enddata
\tablecomments{
Column 1: Central frequency of observation and VLA image.
Column 2: Total bandwidth of each observing frequency.
Column 3: Flux density and its error of the VLA component.
Column 4: Beam major axis in arcsec.
Column 5: Beam minor axis in arcsec.
Column 6: Beam position angle in degrees.
Column 7: RMS of image.
}
\end{deluxetable*}

Our VLA observations utilized broad bandwidths covering the frequency ranges of 2-4 GHz and 8-12 GHz. We determined a two-point spectral index value using the flux densities recovered at the central frequency, 3 and 10 GHz, of each band by calculating
\begin{equation}\label{eqn:alpha}
    \alpha = \frac{\mathrm{log}(S_1/S_2)}{\rm{log}(\nu_1/\nu_2)}.
\end{equation}
The error for the spectral index is then
\begin{equation}\label{eqn:alpha_error}
    \sigma_\alpha = \frac{1}{\rm{ln}(\nu_1/\nu_2)}\sqrt{\left(\frac{\sigma_{S_1}}{S_1}\right)^2 + \left(\frac{\sigma_{S_2}}{S_2}\right)^2}\, ,
\end{equation}
given by standard propagation of errors.

The two-point spectral index of the VLA radio component was found to be $\alpha_{\rm{VLA}} = -0.13 \pm 0.11$, where $S_\nu \propto \nu^\alpha$. A spectral index map is presented in the right panel of Fig.~\ref{fig:vla_intensity}.
Flat spectral index values ($\alpha \geq -0.5)$ are produced by synchrotron self-absorbed radio sources which are actively injecting fresh, relativistic electrons. We take the source compactness and spectral index value as evidence that this radio emission is associated with a radio core. 

\begin{figure}
    \centering
    \includegraphics[scale=0.8]{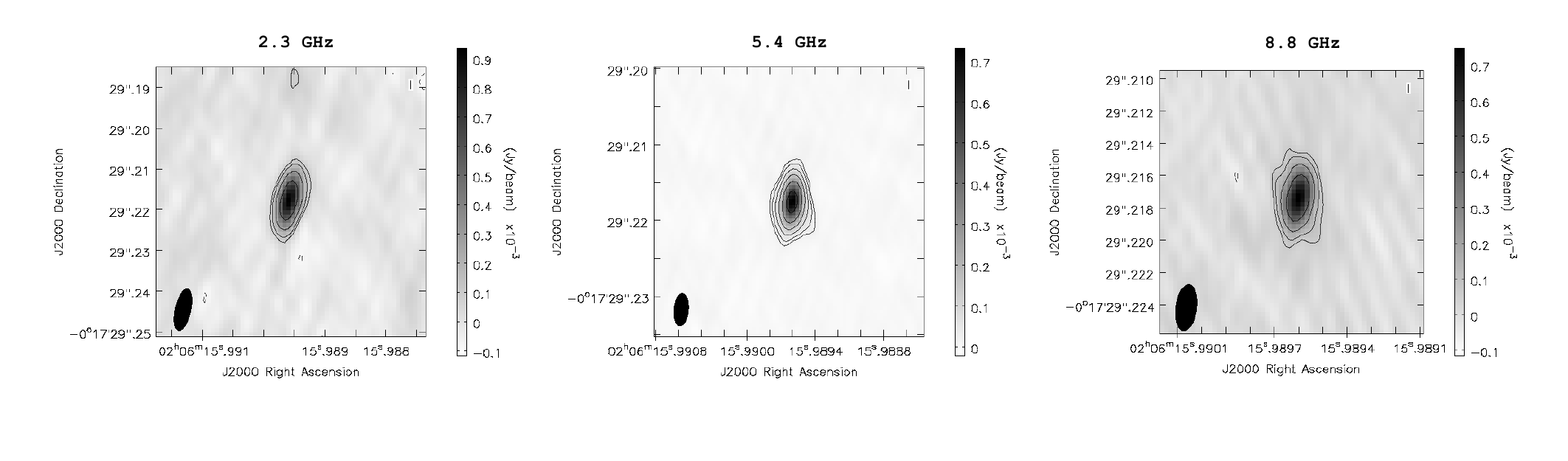}
    \caption{From left to right: 2.3, 5.4, and 8.8 GHz VLBA intensity maps of the radio source associated with Mrk 1018. For the 2.3 and 8.8 GHz image, contours are overlayed at -3, 3, 5, 10, and 20 times the image RMS (27 $\mu$Jy beam$^{-1}$ and 20 $\mu$Jy beam${-1}$, respectively). For the 5.4 GHz image, contours are overlayed at -3, 3, 5, 10, 20, 40, 80 times the image RMS (7 $\mu$Jy beam$^{-1}$) The source is unresolved ($S_{peak}$/$S_{\nu} >$ 0.8) in each image, placing a maximium physical size of 2.4 pc at the redshift of the source, and has a flat spectral index $\alpha= -0.18 \pm 0.04$. Low significance structure is caused by small closure errors in the calibration. The restoring beams are plotted in the lower left hand corner of each image.
    }
    \label{fig:vlba_intensity}
\end{figure}

Fig.~\ref{fig:vlba_intensity} shows the total intensity maps produced by our full VLBA data set on Mrk 1018 at 2.3, 5.4, and 8.8 GHz. Separate images are also produced at 4.3 and 6.6 GHz.
Flux density values were estimated by performing 2D Gaussian fitting to the radio source in each of our VLBA images. The error for each flux density value is given as the sum in quadrature of 10\% of the peak flux value, as is standard for VLBA data, and the RMS of each respective image.
The radio source is unresolved in each image, with the ratio of the peak flux to the total flux density being greater than 0.8 in each case. This places an upper limit on the source size of 2.4 pc.
The low significance (3$\sigma$) structure seen in the contours of the 5.4 and 8.8 GHz images are indicative of small closure errors in the calibration. This is evidenced by the near perfect alignment with the noise structure surrounding the source in these images, though particularly in the 8.8 GHz image. 
After extracting the flux density of the source in the separate 2.3, 4.3, 6.6 and 8.8 GHz images, we performed a non-linear least squares regression analysis to determine the spectral index of the VLBA component. We determined the spectral index of the VLBA component to be $\alpha_{VLBA}= -0.18 \pm 0.04$. Like the unresolved sub-kpc radio emission revealed by our VLA analysis, the unresolved VLBA component also has a flat spectral index value across a similar frequency range. However, approximately half of the VLA flux density at S- and X-band is resolved out when comparing to the VLBA flux density at the same frequencies, indicating emission at larger scales than what is recovered by the synthesized beam of the VLBA.
The flux density values, their associated errors, the restoring beams, and other properties of the VLBA source and images are listed in Table \ref{tab:vlba_obs}. 

\begin{deluxetable*}{ccccccc}[t!]
\tablewidth{0pt}
\tablecaption{Summary of VLBA Observations and Images of Mrk 1018}
\label{tab:vlba_obs}
\tablecolumns{7}
\tablehead{
\colhead{$\nu$ (GHz)} & \colhead{$\Delta \nu$ (MHz)} & \colhead{$S_\nu$ (mJy)} & \colhead{$B_{\rm maj}$ (mas)} &
\colhead{$B_{\rm min}$ (mas)} & \colhead{$PA$ ($\degr$)} & \colhead{$\sigma_{\rm RMS}$ ($\mu$Jy beam$^{-1}$)} \\
\colhead{(1)} & \colhead{(2)} &\colhead{(3)} & \colhead{(4)} & \colhead{(5)} & \colhead{(6)} & \colhead{(7)}
}
\startdata
2.3 & 160 & 1.133 $\pm$ 0.10 & 10.5 & 3.9 & -12.5 & 27 \\
4.3 & 256 & 0.956 $\pm$ 0.08 & 5.5 & 2.4 & -5.6 & 10 \\
6.6 & 256 & 0.921 $\pm$ 0.07 & 3.5 & 1.6 & -4.5 & 15 \\
8.8 & 256 & 0.894 $\pm$ 0.08 & 2.9 & 1.3 & -7.3 & 20 \\
\enddata
\tablecomments{
Column 1: Central frequency of observation and VLBA image.
Column 2: Total bandwidth of each observing frequency. We note that the nominal bandwidth at 2.3 GHz is 256 MHz. However, we removed 3 IFs prior to any calibration due to persistent, strong RFI in these frequency ranges.
Column 3: Flux density and its error of the VLBA component.
Column 4: Beam major axis in arcsec.
Column 5: Beam minor axis in arcsec.
Column 6: Beam position angle in degrees.
Column 7: RMS of image.
}
\end{deluxetable*}

As the source is unresolved, we find the lower limit for the brightness temperature to be 3.8$\times 10^6$ K at 8.8 GHz, well above the thermal limit of active star-forming regions (10$^5$ K; \citealt{Condon_92}). This value is, however, low for typical radio AGN, which often approach the inverse Compton limit of $10^{11}-10^{12}$ K \citep[][]{Kellerman&PT_69}. Similarly, we find the radio power at 8.8 GHz to be $\rm{log}(P_{8.8}) = 21.6$ W Hz$^{-1}$. These values are not dissimilar to the VLBI detections of radio supernova reported by \citet{Smith+98}. However, the detection of only a single unresolved radio component, unlike the multiple unresolved components seen in the northwestern nucleus of Arp 220 by \citet[][]{Smith+98_image}, in addition to the optical emission line BPT classification, provided by the SDSS, as dominated by AGN emission, lead us to believe that the radio emission is indeed associated with a low-power radio AGN, not a radio supernova.

\subsection{Flux Density and Spectral Variability}\label{sec:variability}
It is known that flat-spectrum radio sources exhibit flux density variability over timescales from days to years \citep[][]{Hovatta+07, Chen+13, Sadler+14}. Interstellar scintillation \citep[][]{Walker_98,Said_20} and variations in the Doppler boosting factor \citep[][]{Liodakis+21,Kosogorov+22} are two processes which may cause the observed flux density variability.
For flat-spectrum radio sources, particularly the blazar subclass, this variability is often associated with flaring activity, and is a major source of misidentification of Peaked Spectrum (PS) objects \citep[][]{Kovalev+02,Jauncey+03,Tinti+05,Torniainen+05,Orienti+10}, i.\,e., the radio spectrum becomes peaked where it is expected to be flat. 

\begin{figure}[t!]
    \centering
    \includegraphics[scale=0.7]{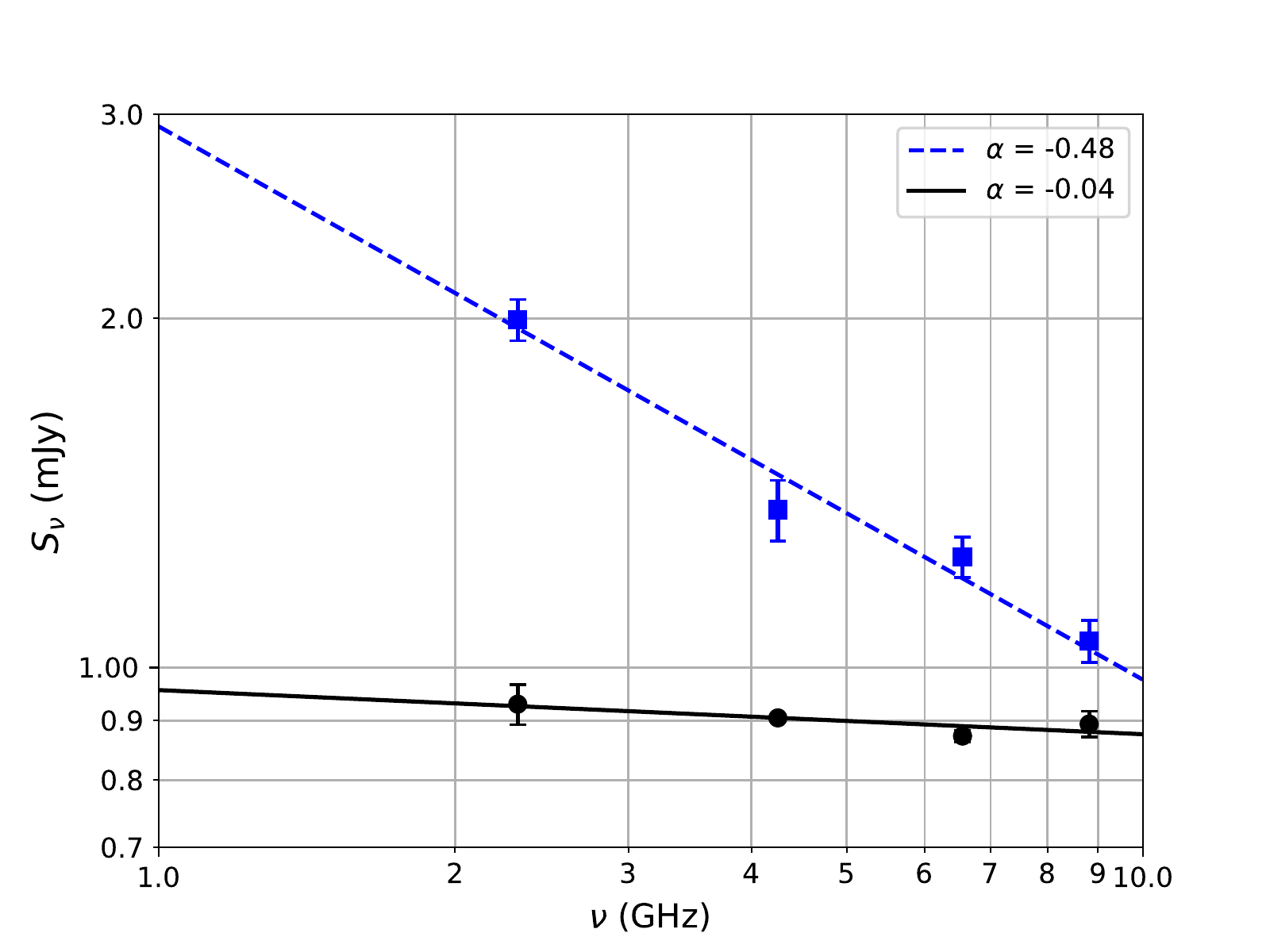}
    \caption{Flux density and spectral index variability of the milliarcsecond-scale radio core associated with Mrk 1018 taken from our multi-epoch, 2.3 - 8.8 GHz Very Long Baseline Array (VLBA) observations. The blue squares and dashed line are from observations taken between August 2014 and June 2016 (VLBA program BS237). The black circles and solid line are from observations between June 2020 and February 2021 (VLBA program BS280). We interpret the spectral change as the result of standard variability associated with a flat-spectrum radio source following a flaring period. We note that the errors shown here minimize the reduced $\chi^2$ of the fit but there is likely a 10\% error in the flux density scale.}
    \label{fig:var_spec}
\end{figure}

We observe such variability of the radio flux densities of Mrk 1018 during our observational campaign, which is shown in Fig.~\ref{fig:var_spec}.
The 2.3 - 8.8 GHz spectral index of the milliarcsecond-scale radio core changes from $\alpha = -0.48 \pm 0.06$ during BS237 (August 2, 2014 to June 21, 2016) to $\alpha = -0.04 \pm 0.02$ during BS280 (June 17, 2020 to February 2, 2021).
This is accompanied by a decrease in flux densities observed more strongly at lower frequencies. The 2.3, 4.3, and 6.6 GHz flux densities decrease by 53\%, 34\%, and 30\%, respectively, between the two programs, whereas the 8.8 GHz flux density does not change within its errors during this same time.
We interpret this as the standard radio variability associated with a flat-spectrum radio source following a flaring period.

If the variability were instead caused by a change in the Doppler boosting factor, we expect to see a systematic change in the observed flux density values, boosted by a factor $\delta^{2+\alpha}$, between the two observing programs. This would be associated with the observed flattening of the radio spectral index, which is indicative of a higher beaming factor. 
However, we do not observe this behavior, instead observing a decrease in flux density values at lower frequencies. 
Additionally, this variability is atypical of what is expected due to adiabatic expansion of a young radio source. During adiabatic expansion, the optically-thin regime will decrease in flux density on short timescales, down to $\sim$10 year \citep[][]{Orienti+10}. However, the optically-thick regime will increase in flux density during this same period, shifting the peak to a lower frequency while maintaining the overall curved shape of the spectrum. Naturally, because the shape of the spectrum has changed entirely, adiabatic expansion is not the mechanism producing the spectral variability. It is likely, then, that the observed change in flux densities for Mrk 1018 between our observing programs is associated with the standard variability of compact radio cores.

\subsection{Core Shift}
We performed multi-frequency positional fitting to test for the core shift phenomenon. 
The radio core, physically the base of the relativistic jet launched by an active SMBH, is located in the region where the surrounding ambient medium has an optical depth $\tau \approx 1$. 
Variation of the optical depth along the jet creates a positional variation of the core with observing frequency, which can be quantified by a power-law scaling relation of the form $r_c \propto \nu^{-1/k_r}$ \citep[][]{Blandford&Konigl_79}. Here, higher frequencies approach the true position of the active SMBH and the jet-launching region itself, as was shown for the case of M87 by \citet[][]{Hada+11}. 
The positions of steep-spectrum features, observed in the optically-thin regime and associated with emission further away from the jet base, are achromatic.
Observing the core shift phenomenon for an unresolved radio source is a simple and powerful way to unambiguously identify the radio core component.

\begin{figure}
    \centering
    \includegraphics[scale=0.7]{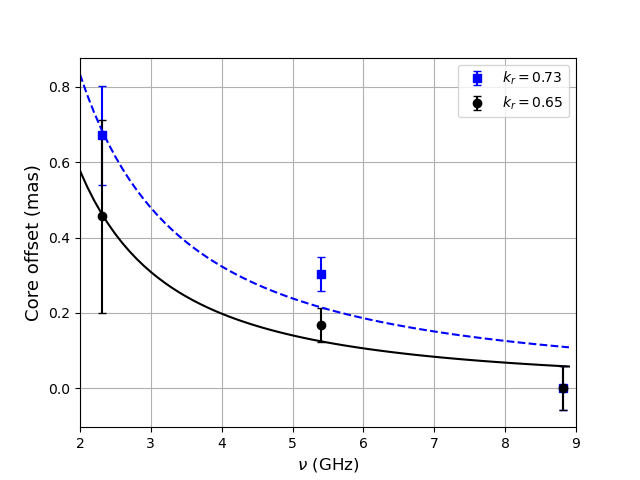}
    \caption{Core shift phenomenon observed for the VLBI component of Mrk 1018. Chromatic variation of the position of radio emission is unique to the radio core, where the opacity of the radio jet is $\tau \approx 1$.
    Blue squares are from observations taken between August 2014 and June 2016 (BS237) with the fit as the dashed, blue line. 
    Black circles are from observations taken between June 2020 and February 2021 (BS280) with the fit as the solid, black line.
    For each observing program, the reference position was taken to be that determined by the respective 8.8 GHz image. 
    Although there is an observed change in the total positional difference of the radio core between the two observing programs, our analysis finds no significant variation of $k_r$ over this same period. 
    }
    \label{fig:core_shift}
\end{figure}

Fig.~\ref{fig:core_shift} shows the result of our core shift analysis for our two observing programs. We observe the expected frequency-dependent positional variations associated with a radio core for Mrk 1018.
The reference position for each program was taken to be the position determined by fitting the unresolved radio component in the two 8.8 GHz images separately. Angular offsets for the 2.3 and 5.4 GHz positions were determined for each program using the respective 8.8 GHz fitted position.
For BS237, this corresponds to a positional shift of 0.67 mas; for BS280, 0.46 mas.
The value of $k_r$, which is physically motivated by the electron energy spectrum, and the magnetic field and particle density distributions of the jet \citep[][]{Lobanov_98}, was found to be $0.73\pm0.34$ for BS237, and $0.65\pm0.24$ for BS280. Although there is an observed change in the total positional difference of the radio core between the two programs, our analysis finds no significant variation of $k_r$ over this same period. Implications for relativistic jet models or the ambient medium hosting the radio jet are outside the scope of this paper. We report only the chromatic variation of the position of the radio emission, further establishing this source as a radio core.

\subsection{Proper Motion with the VLBA}\label{sec:proper_motion}
As proposed by \citet{kim+18}, the changing-look nature of Mrk 1018 could be explained by a gravitational wave recoil. In this scenario, two or more individual SMBHs merged and, due to anisotropic emission of low-frequency gravitational waves before merger, the newly-merged SMBH experienced rapid acceleration, causing it to leave the host galaxy's gravitational center. 
As the recoiling SMBH (rSMBH) oscillates about the center of mass of the host galaxy, its accretion disk is tidally disrupted as the rSMBH passes through pericenter in its orbit. These accretion disk disruptions from tidal forces are able to explain both the spectral Type variability of the AGN as well as its observed optical and X-ray flux variability \citep[][]{husemann+16,lyu+21}. 

Our observational program offers the unique ability to test the rSMBH hypothesis using multi-epoch, high resolution radio observations with the VLBA. Our observations were designed to search for supermassive black hole binaries (SMBHBs) and evidence of recoiling SMBHs in a sample of six late-stage galaxy mergers (Walsh et al. in prep). 
For the proper motion analysis here, we used X-band observations, centered at 8.8 GHz observing frequency, to obtain the highest astrometric precision ($\sigma \propto \theta/(S/N)$) for the position of the radio AGN associated with Mrk 1018. We also considered the 8.8 GHz data most reliable due to the flux density variability of the radio core at 2.3, 4.3, and 6.6 GHz, as mentioned previously (see Sec.~\ref{sec:variability}). 
Unresolved source structure can affect the positional fitting of the radio core by the presence of non-core ejecta, such as a jet knot.
By excluding the $\nu <$ 8 GHz data from our proper motion analysis, we attempted to mitigate any biases introduced by possible source sub-structure and its potential variability.

To monitor the position of Mrk 1018 with higher temporal sampling than only the average position in each of the BS237 and BS280 programs, we combined observations within each program such that an image produced from each independent data set would reach a detection threshold of at least 10$\sigma$. This corresponded to a net integration time of 3 hours for the observations taken during BS237 (Aug 2014 - June 2016), and 2 hours for observations taken during BS280 (June 2020 - Feb 2021).
The difference in integration time is due to the change in peak brightness of the radio source, which was found to have increased at 8.8 GHz during the 4-year gap in our observing program.

\begin{figure}
    \centering
    \includegraphics[scale=1]{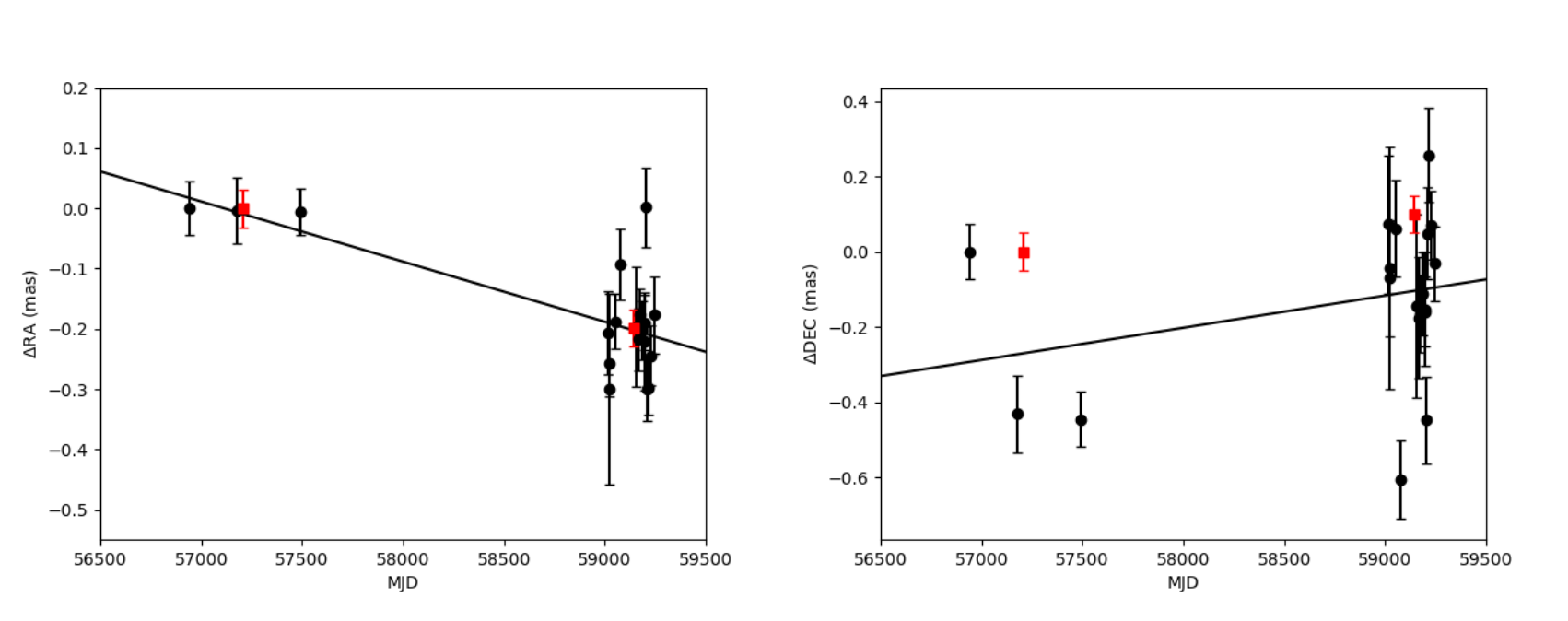}
    \caption{\textit{Left}: Angular offsets in Right Ascension of Mrk 1018's milliarcsecond radio core as a function of MJD. The best-fit line represents a proper motion of -36.4 $\pm$ 8.6 $\mu$as yr$^{-1}$, corresponding to a significance of 4.2$\sigma$. Positions were determined using the 8.8 GHz data.
    \textit{Right}: Angular offsets in Declination of Mrk 1018's milliarcsecond radio core as a function of MJD. The best-fit line represents a proper motion of 31.3 $\pm$ 25.1 $\mu$as yr$^{-1}$ (1.2$\sigma$). Positions were determined using the 8.8 GHz data.
    For each, the reference position was taken to be that determined at MJD 56940. Errors are $1\sigma$ formal uncertainties.
    Black circles represent the 3- and 2-hour integration images for the two observation programs: BS237 and BS280, respectively. 
    Red squares represent the total concatenated data for each program. 
    }
    \label{fig:proper_motion}
\end{figure}

For each image, the position was derived by fitting a 2D elliptical Gaussian to the radio core. 
To confirm the results of our image plane fitting, we also fit the visibility data directly for a small number of data sets. We found an agreement in the position of the radio core between the image plane fitting and visibility fitting procedures to $<0.001$\%. However, position uncertainty is difficult to quantify from visibility fitting, whereas the root mean square of the image can be measured directly (see Section 2.4 of \citet{Deller+19} for additional discussion). Because of this, we proceeded by using image plane fitting to perform our proper motion analysis.
The reference position was taken to be that determined at MJD 56940. Angular offsets were found by taking the difference between the position at a later MJD from the reference position. 
A least squares linear regression was performed to search for proper motion of the radio core. Our analysis found a proper motion of -36.4 $\pm$ 8.6 $\mu$as yr$^{-1}$, corresponding to 4.2$\sigma$, in Right Ascension for the radio core associated with Mrk 1018. In physical units, this is $0.1c \pm 0.02c$ at the redshift of the source.
We report no significant proper motion of the radio core in Declination (31.3 $\pm$ 25.1 $\mu$as yr$^{-1}$). The results of our proper motion analysis are shown in Fig.~\ref{fig:proper_motion}. We discuss the significance and potential causes of this proper motion in Sec.~\ref{sec:discussion}.

\section{Discussion} \label{sec:discussion}
The radio properties of Mrk 1018 indicate that it is a low-luminosity, core-dominated, flat-spectrum radio source. 
Using multi-epoch, high frequency VLBA observations, we observed proper motion of Mrk 1018's radio core with an apparent velocity of $0.10c \pm 0.02c$.
In this section, we discuss the possible physical mechanisms that could produce this observed proper motion. We first discuss potential external effects which could falsely contribute to the observed proper motion of Mrk 1018. Then, we propose and discuss 3 different astrophysical scenarios which could cause the observed mobility of the radio AGN in Mrk 1018: secular evolution of the radio brightness; the orbital motion from a SMBHB; and a gravitational wave recoil.

\subsection{External Effects}\label{sec:systematics}
The most important external effect which could falsely contribute to the observed proper motion of Mrk 1018 is any observed variation in the position of the primary phase reference calibrator, C1. As mentioned in Sec.~\ref{sec:vlba_obs}, the phase referencing technique sets the coordinates of the weak target source by referencing it to the phase calibrator. When performing differential astrometry, the positional difference between the phase reference calibrator and the target source is measured. Thus, any observed variation in the position of the phase reference calibrator will contribute to a potential proper motion of the target source. 

\begin{figure}
    \centering
    \includegraphics[scale=0.9]{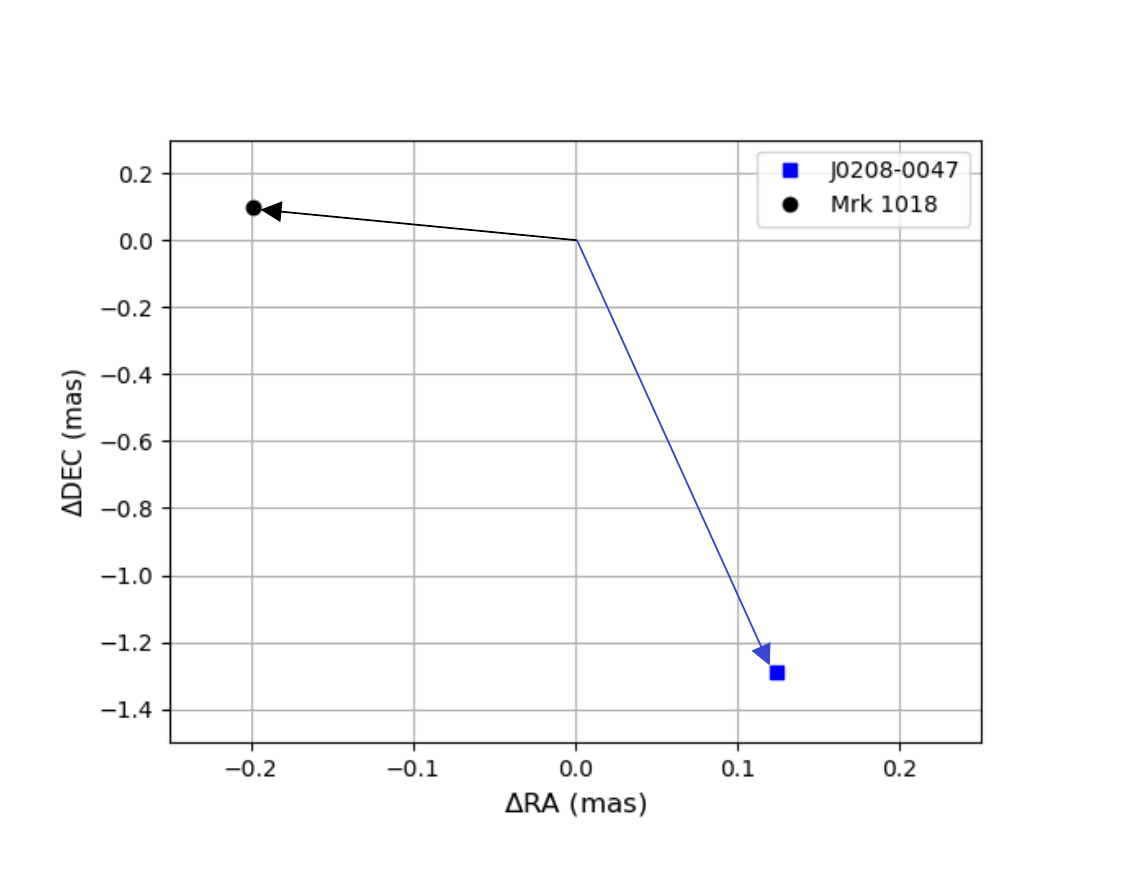}
    \caption{Positional displacement vectors for Mrk 1018 and the primary phase reference calibrator J0208-0047 (C1). For C1, the positional difference is that between its observed position during the BS237 program from its position given in the ICRF3 \citep[][]{charlot+20}, which matches its observed position in the later BS280 program. It is clear that the observed proper motion of Mrk 1018 is uncorrelated with any positional variation of C1.}
    \label{fig:phcal_shift}
\end{figure}

As shown in Fig.~\ref{fig:phcal_shift}, the observed position of C1 changed significantly over the duration of our observing campaign.
Most likely, this can be attributed to a change in the correlation position of C1 due to the adoption of the ICRF3 in between our two observing programs.
C1's ICRF3 position differs from its ICRF2 position by 0.196 mas in RA and -0.892 mas in DEC. In both realizations of the ICRF, the positional difference we measure is consistent with the ICRF difference within measurement errors. 
If this difference in correlation positions is not accounted for, we would introduce this same systematic effect onto the positions of Mrk 1018.
By forcing the correlation position of C1 to match its ICRF3 value for both observing programs, we are ensuring that this does not affect the astrometry of Mrk 1018. 

In support of this argument is the magnitude and direction of the change in position of C1 and observed proper motion of Mrk 1018. If the change in position of C1 were contributing to the proper motion of Mrk 1018, the magnitude and direction of proper motion for Mrk 1018 should reflect what is observed for C1, in addition to potential real proper motion of Mrk 1018. After correcting for the difference in correlation position of C1, the magnitude and direction of proper motion for Mrk 1018 is distinct from what is observed for C1. Again, this demonstrates that the change in correlation position of C1 related to the adoption of the ICRF3 is not affecting our astrometric analysis of Mrk 1018.

Because our observations of Mrk 1018 were part of a larger campaign, we can utilize the multi-source nature of our observations to search for more potential contributions from external effects. If there were systematic errors occurring with the VLBA, such as ubiquitous pointing errors, we would observe the exact same difference in observed versus ICRF3 positions for all of the phase calibrators in our campaign. For the other five primary phase reference calibrators used in our observing campaign (Walsh et al. in prep), the observed positional difference is unique to each calibrator. We are confident, then, that our calibration procedure has accounted for artifacts which may contribute to false proper motion of Mrk 1018. 

\subsection{Secular Evolution}
The observed proper motion of the radio core in Mrk 1018 may be caused by secular evolution of the radio source. There are a number of source-intrinsic or geometric effects which may contribute to the positional variations observed in the radio core. These can include, but are not limited to: jet Lorentz factor, electron density, jet viewing angle, and jet precession \citep[see][]{Moor+11}. 
For unresolved sources, these phenomenon manifest themselves as changes in the brightness distribution of the source.. 

Generally, studies on the proper motion of VLBI sources
have determined that brightness evolution is the physical motivation for the observed motion.
\citet[][]{Fomalont+11} observed four VLBI phase reference sources at 8.6, 23, and 43 GHz over a one year period to assess the stability of their radio cores and the impact of structure effects on positions in the ICRF. Although each source was found to be compact at 8.6 GHz, all had multiple components at 43 GHz, which could confound any astrometric analyses at 8.6 GHz. The radio cores were found to be stationary over the observing period, though two sources had detectable motion in the direction of their jet axis. 
\citet[][]{Moor+11} found a general correlation between the direction of proper motion and the characteristic direction of the VLBI radio jet for 62 VLBI sources with a core+jet morphology, implying that observed proper motion of VLBI sources is in general associated with source brightness evolution. 
Notably, they found that most of their sources had proper motion values within the range of 0.01 - 0.1 mas yr$^{-1}$, which contains the observed proper motion value of Mrk 1018's radio core at the lower end of this range.
\citet[][]{Titov+22} examined four radio AGN which showed shifts in their positions by 20-130 mas over a two decade period. Each showed drastic changes in their brightness distribution over this period, linking the observed positional shifts to intrinsic source evolution. 

For each of these studies, the sample consisted of complex VLBI radio sources which displayed more than one component. 
Our VLBA images show no emission components other than an unresolved radio core. It is unreasonable, then, to define a characteristic direction for any radio outflow components which may be ejected from the core. That is to say, we cannot perform an analysis comparable to \citet[][]{Moor+11} to assess the influence of the characteristic jet direction on the direction of proper motion for this source. Nor do we observe the drastic angular differences most likely associated with superluminal motion of jet components reported by \citet[][]{Titov+22}. We have attempted to mitigate any biases introduced from sub-resolution structure \citep[e.g.,][]{Fomalont+11} by selecting the highest frequency from our observations which showed the most stable flux density across the length of our observing campaign. However, we cannot justifiably rule out intrinsic brightness evolution as the dominant means producing the observed proper motion. Further observations at higher frequency and/or longer baselines may be able to rule out some of these scenarios.

Similarly, precession of the radio jet axis has been identified in a number of VLBI sources \citep[][]{Stirling+03,Marti-Vidal+11,Britzen+17,Algaba+19,Li+20}. It is important to note that compelling evidence of jet precession is mostly realized by the complex morphologies observed in these VLBI systems. Position angle changes of the VLBI jet axis are perhaps the most commonly observed characteristic of precession \citep[][]{Lobanov+05,Caproni+13,Kun+14,Dey+21}. Often, precession is observed morphologically as a sinusoidal ridge line of jet knots in intensity maps of the milliarcsecond jet, which is the product of Doppler beaming within the helical jet structure \citep[][]{Worrall+07,An+10,Kharb+19}. In other cases, long-term sinusoidal variations in the position of the peak brightness are presented as the manifestations of jet precession \citep[e.g., 3C 66B:][]{Sudou+03}. The single component morphology of Mrk 1018 prevents us from testing for jet precession via methods involving extended structures, e.\,g, position angle and ridge line analyses. Though flux density and positional variations of the radio core do not show clear indication of periodic behavior, follow-up observations are required to robustly test this scenario. As such, radio jet precession remains a plausible explanation for the observed proper motion of Mrk 1018's radio core.

\subsection{SMBH Binary}
\label{sec:binary}
Mrk 1018 is hosted by a late-stage galaxy merger, as is evident by the tidal deformities present in the available optical images of the galaxy coupled with the presence of only a single stellar core. Late-stage galaxy mergers are ideal laboratories to search for supermassive black hole binaries (SMBHBs), the progenitors of low-frequency gravitational waves, because the resident SMBHs deposited by the galaxy merger should have evolved significantly due to dynamical friction. VLBI is a powerful observational tool to search for binary systems in local ($z\leq0.1$) galaxy mergers, in particular, due to its capability to resolve the individual binary constituents at orbital separations down to $r\leq10$ pc \citep[][]{Burke-Spolaor+18}. The detection of dual, flat-spectrum radio cores separated by 7.3 pc using VLBI has so far identified the most confidently evidenced SMBHB to date \citep[][]{Rodriguez+06}.
Our multi-frequency VLBA campaign targeting Mrk 1018 provides the deep sensitivities required to probe the possible binary nature of the resident SMBHs.

We created images of $1\arcsec\times1\arcsec$ in size, or 842 pc $\times$ 842 pc at the redshift of Mrk 1018, using our VLBA data at 2.3, 5.4, and 8.8 GHz using the full data set of our observing programs on Mrk 1018. The 5.4 GHz image is the concatenation of the two C-band frequency ranges, centered at 4.3 and 6.6 GHz. We chose to do this to reach the deepest sensitivity at C-band, effectively doubling the bandwidth at this frequency. Each of our VLBA images fully encapsulates the single radio component associated with Mrk 1018 that was revealed by our VLA analysis.
We then searched each image for a secondary radio companion, setting the detection threshold to 5$\sigma$ in each image, corresponding to a peak flux density limit of 135, 35, and 100 uJy beam$^{-1}$ for the secondary companion in the 2.3, 5.4, and 8.8 GHz images, respectively. We find no evidence for a radio-emitting companion in any of the images. Assuming the secondary SMBH is radio emitting, this could be caused by two possibilities: our observations are not sensitive to the radio-faint secondary companion, or the orbital separation of the binary constituents is smaller than the angular scales probed by our observations. 

To test our sensitivity to a potential secondary radio source, we consider the likelihood of detecting a secondary AGN given the depth of our imaging. The $5\sigma$ luminosity limit on a compact source probed by each of the three images is 5.7, 1.5, and 4.3$\times10^{20}$ W Hz$^{-1}$ for the 2.3, 5.4, and 8.8 GHz images, respectively. 
The full bandwidth, C-band observations set the most constraining limit. Given the deep limit set here, the observation should have certainly detected a radio-loud AGN, and also detected the bulk of radio-quiet AGN, based on their luminosity functions \citep[e.g.][]{Padovani+09}. Thus, if there are two AGN in this system, we should have detected the second. However, we cannot place meaningful limits on a secondary ``naked'' SMBH that is not accreting.

The second possibility is that the orbital separation of the SMBHB is smaller than the highest angular resolution of our VLBA observations. For the case of Mrk 1018, the binary's orbit would need to be $<$ 2.4 pc, as this is the physical scale of the beam size which produced the 8.8 GHz image (Fig.~\ref{fig:vlba_intensity}) and there is only a single, unresolved source in the image, which would not be the case for even a marginally resolved SMBHB orbit. We cannot rule this scenario out with our VLBA observations alone, however, as the positional variations we observe could indeed be consistent with 
the orbital motion of binary constituents which are separated by a distance smaller than the physical scales probed by the 8.8 GHz image alone (e.\,g., 3C 66B: \citealp{Sudou+03},  J1918+4937: \citealp{Hu+20}). However, K18 provide a compelling argument against the SMBHB hypothesis using the H$\alpha$ variability.
If the binary is of unequal mass ratio, the broad-line component, red or blue wing, with the higher velocity offset will always be of smaller width, due to the broad emission line width scaling as $M^{1/2}$. This behavior is not observed in the variability of the broad H$\alpha$ line. 
Additionally, \citet[][]{Hutsemekers+20} find no evidence for SMBHB evolution in spectropolarimetric observations of Mrk 1018.
While we cannot directly rule out the SMBHB hypothesis from our radio observations, when considering the optical emission line properties, it is unlikely that Mrk 1018 hosts a SMBHB which gives rise to the proper motion we have observed.

\subsection{Recoiling SMBH}
In a rSMBH scenario for this source, the variable red- and blue-wing velocity offsets observed in archival optical spectra of Mrk 1018 represent the oscillation of the rSMBH, carrying with it two kinematically distinct broad-line regions \citep[][]{kim+18}, about the center of mass of the host galaxy. Due to tidal disruption of the accretion disk as the rSMBH passes through pericenter in its orbit, this scenario has been adopted as an explanation of the changing-look nature of this AGN. By modelling the time variability of the H$\alpha$ red- and blue-wing velocity offsets, K18 derived orbital parameters for the presumptive rSMBH. They found the rSMBH to be on an eccentric orbit with eccentricity $\epsilon=0.94$, orbital period $P=29.2$ yr, and a semi-major axis $a=0.004$ pc, though this value aws determined by fixing the orbital inclination angle to match the inclination angle of the host galaxy (49\degr; \citealp{peng+10}).

We begin by discussing difficulties of the K18 model and the physical quantities derived from our observations. 
If the K18 model accurately recovers the orbital parameters of the rSMBH, we should not observe any proper motion of the radio core with our observations because the semi-major axis of the orbit is too small to be probed by our observations. This is shown via the following considerations. At 8.8 GHz, the restoring beam for the VLBA is about 1 mas, and is approximately such for our 8.8 GHz images. We required that each of the 8.8 GHz images produced for the astrometric analysis reach a minimum S/N of 10. 
Then, the smallest error achievable for the astrometric precision of each localization, ignoring contributions from phase referencing errors, is approximately $\theta/\left(2*S/N\right)$ or $\sim 50\, \mu$as, which is 0.042 pc at the redshift of Mrk 1018. This is still a factor of 10x higher than the semi-major axis of the rSMBH orbit predicted by K18 (0.004 pc).
Additionally, our observing campaign observed only $\sim 25\%$ of the K18 inferred orbit. Considering the trajectory of the orbit, this could correspond to pericenter or apocenter, when minimal proper motion would be observed. Certainly, this, combined with the drastic difference in scale size between the inferred orbit and that probed by our observations, should inhibit our ability to detect proper motion of the proposed rSMBH.
However, we do observe proper motion in this source, corresponding to a total change in position of 0.165 pc. This is a factor of 40x greater than the K18-derived semi-major axis. Though there is uncertainty in the inclination angle of the rSMBH orbit, the observed change in position is certainly too great for inclination effects to account for alone.

Furthermore, assuming the orbit of the rSMBH is Keplerian, which is valid given the non-relativistic velocities of rSMBHs, Kepler's third law can be used to recover the mass contained within the orbit of the rSMBH. We take the SMBH mass of Mrk 1018 to be $\mathrm{log}(M_{BH}/M_\odot) = 7.84$ \citep[][]{Noda&Done_18} and use the orbital parameters derived by K18 to find this mass. Then, we find the required mass contained within the orbit to be unphysical since it is negative in this model. This alone raises serious questions about the plausibility of the K18 model. To make this a physically plausible model would require the inclusion of a second SMBH. The simplest model in this case is one in which the second SMBH mass is significantly greater than that of Mrk 1018. In this scenario, Mrk 1018 then has easier access to the gas in the circumbinary disc and thus accretion, which is, naturally, required to explain the changing-look nature of this AGN. Assuming a mass ratio of $q=M_2/M_1=0.1$, the inferred mass contained within the orbit of the rSMBH now becomes $\sim 10^9 M_\sun$. If this mass were of a purely stellar population, the stellar mass surface density $\Sigma$ bounded by the area of the orbit is $\sim 4\times10^{11}$ M$_\sun$ pc$^{-2}$. \citet[][]{Hopkins+10} found that the stellar mass surface density for any dense stellar system maximizes at $\Sigma_{\rm{max}} \sim 10^5$ M$_\sun$ pc$^{-2}$, 6 orders of magnitude lower than this inferred mass. Thus, a second SMBH must be used to explain this inferred mass. As discussed in Sec.~\ref{sec:binary}, we cannot rule this scenario out directly with our radio observations, though the H$\alpha$ variability \citep[][]{kim+18} and absence of spectropolarimetric signatures \citep[][]{Hutsemekers+20} provide evidence against the presence of a SMBHB.

We can further argue the proper motion is not due to a recoil scenario because the observed proper motion of the radio core has an apparent velocity of $0.1c$. This is significantly higher, by a factor of 6, than the theoretical predictions for the maximum recoil velocity of rSMBHs (5000 km s$^{-1}$). 
Nominally, radio cores do not experience superluminal motion because the core represents a stationary point in the jet where the jet opacity reaches $\tau \approx 1$. rSMBHs present a unique situation, however.
If the radio jet of the rSMBH is slightly misaligned with the observing axis, such as for a blazar, it is possible to induce a velocity-boost effect similar to that seen in the superluminal motion of knots flowing along the jet axis. One can calculate this boost as

\begin{equation}\label{eqn:sprlum_mtn}
    \beta_{obs} = \frac{\beta sin(\phi)}{1-\beta cos(\phi)},
\end{equation}
where $\beta_{obs}$ is the observed apparent velocity in units $v/c$, $\beta$ is the actual velocity, and $\phi$ is the viewing angle to the jet axis. 
We can test whether a radio core associated with a rSMBH can reach an apparent velocity of 0.1$c$ by assuming the rSMBH travels at 5000 km s$^{-1}$ ( $\beta = 0.0167$), the maximum recoil velocity as determined by simulations, and varying the viewing angle to the jet axis. We found that even for this maximum recoil velocity, no value of $\phi$ could yield a $\beta_{obs}$ of 0.1, corresponding to the factor of 6 required to match the maximum recoil velocity to the observed apparent velocity. This assures that viewing angle effects are not creating an improbably high recoil velocity.

It is challenging to reconcile the differences between our analyses and the K18 rSMBH parameters by only considering a rSMBH. Though we have attempted to mitigate biases from sub-resolution structure by selecting the highest frequency in our observations, which is less affected by steep-spectrum jet components, a rSMBH alone is not sufficient to explain the observed proper motion. It is likely that the observed proper motion is dominated by an unresolved jet component. This does not rule out the rSMBH hypothesis, however. Follow-up observations of this source are needed to confirm the nature of the proper motion. If there is a significant contribution to the observed proper motion from the orbital motion of a rSMBH, the radio core should exhibit periodic, non-linear motion over an extended time baseline. This would provide the most compelling evidence that Mrk 1018 is a rSMBH. 
Multi-epoch, long-integration observations with VLBI at higher frequency will mitigate contributions from sub-resolution structure, and provide higher astrometric precision than the 8.8 GHz observations used in this paper. Observations with space VLBI or the Event Horizon Telescope (EHT) may offer even more powerful means to identify any proper motion in the radio core associated with a rSMBH due to the longer baselines and, certainly for EHT, higher frequency observations.

\section{Conclusions}
\label{sec:conclusions}
Mrk 1018 is a known changing-look AGN which has oscillated between Type 1.9 and Type 1 over 40 years. \citet[][]{kim+18} proposed the time-varying flux and spectral characteristics of this AGN as the signatures of a recoiling SMBH. 
We have obtained multi-frequency, multi-epoch observations of Mrk 1018 with the Very Long Baseline Array during a 6.5-year observational campaign. These observations revealed a compact ($\leq$ 2.4 pc), low luminosity (${\mathrm{log}(P_{8.8})} = 21.6$ W Hz$^{-1}$), synchrotron-dominated ($T_b = 3.8\times10^6$ K) radio core associated with this changing-look AGN. 
 
Using our 8.8 GHz data, we found a proper motion of -36.4 $\pm$ 8.6 $\mu$as yr$^{-1}$, a significance of 4.2$\sigma$, in RA of the radio core. No significant proper motion (31.3 $\pm$ 25.1 $\mu$as yr$^{-1}$) was found in DEC. At the redshift of the source, this is an apparent velocity of $0.1c \pm 0.02c$ and corresponds to a physical displacement of 0.165 pc during the length of our observing campaign. After ruling out external effects such as correlation position differences of the phase reference calibrator and instrumental errors, we discussed the proper motion as potentially arising from three physical mechanisms: secular evolution of the brightness distribution; SMBH binary; or rSMBH.

We were not able to rule out the possibility that the proper motion is caused by secular evolution of the brightness distribution of the source, due to either source-intrinsic or geometrical evolution, alone.
The deep sensitivities attained by our observing campaign make the presence of a secondary, radio-faint AGN companion, thus the SMBHB scenario, unlikely, down to limiting radio luminosities of 5.7, 1.5, and 4.3$\times10^{20}$ W Hz$^{-1}$ at the 2.3, 5.4, and 8.8 GHz. 
For a rSMBH scenario, the apparent velocity of the radio core's motion is too great to be explained purely by the motion of a rSMBH, which theoretically maximizes at 5000 km s$^{-1}$ (0.0165$c$). This is true even when considering viewing angle effects, which could lead to an artificially high apparent velocities. 
In addition, the K18 model predicts an unphysical mass contained within the hypothesized rSMBH orbit given the SMBH mass estimate of Mrk 1018 \citep[][]{Noda&Done_18}, and is too small to be probed even by our VLBI observations.

While the rSMBH hypothesis remains a viable option to explain the changing-look nature of the AGN, our VLBA observations favor a scenario in which the proper motion we observed of the radio core is dominated by an outflowing component of the radio jet.
Future observations of this source must either be performed at higher frequencies, with significantly longer baselines, or a combination of these to both probe the hypothesized orbit accurately and to minimize potential contributions from sub-resolution jet components to the positional fitting of the radio core. Nevertheless, the rSMBH hypothesis remains an intriguing method to explain short-term AGN variability and Mrk 1018 remains an exciting candidate for future follow-ups on the still-elusive recoiling SMBH.  

\section*{Acknowledgements}
We thank the referee for comments that have helped improve the manuscript. GW thanks Mark Reid for the helpful discussions involving the calibration and astrometry procedures. GW was supported in this work by NASA Grant \#80NSSC20M0055. GW and SBS were supported in this work by NSF award grant \#1815664.  The National Radio Astronomy Observatory is a facility of the National Science Foundation operated under cooperative agreement by Associated Universities, Inc. Part of this research was carried out at the Jet Propulsion Laboratory, California Institute of Technology, under a contract with the National Aeronautics and Space Administration.The BS237 observing campaign was initiated by the listed authors and collaborators L.~Petrov, A.~Deller, J. McKean, J. Moldeon, and M. Lee. The NANOGrav collaboration, which funded some components associated with this research, receives support from National Science Foundation (NSF) Physics Frontiers Center award \#1430284 and \#2020265. This research has made use of NASA's Astrophysics Data System Bibliographic Services.

\facilities{VLA, VLBA}
\software{AIPS \citet[][]{aips_99},
CASA \citet[][]{CASA_22},
numpy \citet[][]{numpy+11},
scipy \citet[][]{scipy+20},
matplotlib \citet[][]{Hunter_07}
}

\bibliography{bib}{}
\bibliographystyle{aasjournal}

\end{document}